# Fully Consistent Density Functional Theory Determination of the Insulator-Metal Transition Boundary in Warm Dense Hydrogen

Joshua Hinz[1], Valentin V. Karasiev[1*], S. X. Hu[1], Mohamed Zaghoo[1], Daniel Mejía-Rodríguez[2],

S.B. Trickey[2], and L. Calderín[3]

*vkarasev@lle.rochester.edu

[1]Laboratory for Laser Energetics, University of Rochester, Rochester, NY 14623

[2] Quantum Theory Project, Department of Physics, University of Florida, Gainesville, FL 32611

[3]Department of Materials Science and Engineering, University of Arizona, Tucson AZ 85721

Last Revised: 7 Feb. 2020

Using conceptually and procedurally consistent density functional theory (DFT) calculations with an advanced meta-GGA exchange-correlation functional in ab initio molecular dynamics simulations, we determine the insulator-metal transition (IMT) of warm dense fluid hydrogen over the pressure range 50 to 300 GPa. Inclusion of nuclear quantum effects via path-integral molecular dynamics (PIMD) sharpens the metallic transition and lowers the transition temperature relative to results from Born-Oppenheimer (BO) MD. BOMD itself gives improved agreement with experimental results compared to previous DFT predictions. Examination of the ionic pair correlation function in the context of the abrupt conductivity increase at the transition confirms a metallic transition due to the dissociation of molecular hydrogen that coincides with an abrupt band gap closure. Direct comparison of the PIMD and BOMD results clearly demonstrates an isotope effect on the IMT. Distinct from stochastic simulations, these results do not depend upon any ad hoc combination of ground-state and finite-T methodologies.

*Motivation* The liquid–liquid insulator-to-metal transition (IMT) of warm dense hydrogen isotopes is a crucial phenomenon for giant planet structure and dynamics. Because hydrogen has the highest relative abundance in the universe, accurate determination of the IMT is key to modelling the interior dynamics and evolution of these Jovian-like planets [1]. The metallization of hydrogen is believed to be the catalyst for H–He de-mixing, a central feature in internal structure models that results in a clearly demarked boundary between He-rich and He-poor layers. Additionally, the conductive behavior of dense hydrogen is crucial to understanding the dynamo and the depth of zonal flow in such planets [2]. The recent Juno space mission measurements of Jupiter's magnetic fields [3] have heightened scrutiny of those problems [1]. An essential pre-requisite for quantitative modeling of all these planetary phenomena is an accurate equation of state (EOS) that describes the onset and character of the IMT correctly [4]. Independently of planetary physics, an accurate EOS for hydrogen and its isotopes also is essential for progress in inertial confinement fusion research [5].

Despite its importance, accurate determination of the IMT boundary remains an experimental and theoretical/computational challenge. In this letter we make a major step forward on the theory/computation side by providing a single, conceptually consistent density functional theory (DFT) treatment, with the best-balanced modern approximate exchange-correlation functional available. We provide results both with and without nuclear quantum effects (NQEs) that are mostly consistent with experimental findings and with best available combinations of DFT and stochastic methods.

*State of the Art.* Beginning with experiments, gas-gun experiments of Weir *et al.* [6] that measured the resistivity of deuterium but not directly the temperature, placed an IMT boundary point at 140 GPa with an estimated temperature $T_m > 2500$ K. Static compression experiments (SCEs) on hydrogen [7-9] determined a set of IMT boundary points, as interpreted by plateaus in the laser heating curve, to occur at pressures $P_m$ <180 GPa and $T_m$ significantly below the Weir *et al.* value. The optical measurements of Ref. 8 indicated that the plateaus in the laser heating curve immediately precede a substantial increase in the reflectivity. This suggests that such plateaus are evidence of the latent heat of a first-order transition, but see further detail below. Follow up SCEs by Zaghoo and Silvera [10] resolved the measured reflectance of hydrogen further, with the result being an upward shift of the IMT boundary as compared to previous SCEs by ≈500 K, but still significantly lower than the Weir *et al.* results. However, McWilliam and coworkers [11] later determined that the laser heating curve plateaus resulted not from the latent heat of the IMT but from onset of hydrogen absorbance at the optical frequencies used in the experiments of references [7-9].

Subsequent SCEs by Zaghoo [12] on deuterium at the same pressures as their previous work on hydrogen (9) shifted the IMT boundary to a value ≈600 K higher than for hydrogen. Identification of that upward shift as an isotope effect was strengthened by the fact that the temperature shift is on the order of the bond energy difference between diatomic hydrogen and diatomic deuterium. The IMT boundary location was clouded further, however, by the dynamic compression experiments (DCEs) of Knudson *et al.* [13] on deuterium which placed the IMT at unprecedented pressures, $P_m$ > 300 GPa, with weak T-dependence. A reanalysis of the Knudson data by Celliers (see supplemental material of Ref. [14]) reduced the estimate to $T_m$ < 1000 K. That reanalysis, in conjunction with the DCE results of Celliers [14], forms a clear IMT boundary for deuterium and further supports the notion of differing boundaries for hydrogen and its isotopes as can be seen in the present state of affairs displayed in Figure 1.

However, there is still a debate on the experimental side pertaining to Celliers' reanalysis of the Knudson *et al.* DCEs, results, see reference [15,16]

Theoretical predictions of the IMT boundary have been broadly inconsistent, agreeing with some, but not all, of the experimental data. The obvious causal suspect is the diversity of approximations within calculations. DFT, the most-used tool, is, in principle, an exact mean-field solution to the quantum many-body electronic problem. However, the exchange-correlation (XC) free-energy functional must be approximated [17,18]. Some of the earlier hydrogen IMT predictions from DFT were done in the Born-Oppenheimer (BO) approximation [19-21] and employed the ground-state Perdew–Burke–Ernzerhof (PBE) generalized gradient approximation (GGA) XC functional [22]. Those results demonstrated no more than $\approx$ 25 GPa pressure difference from earlier SCEs [7-9] and indicated an IMT phase boundary occurring in the lower-pressure liquid regime. (Note that use of a ground-state XC functional with Fermi-Dirac finite-T occupations rather than a full free-energy functional is well justified in the temperature range of interest [23,24]). With the addition of nuclear quantum effects (NQEs) [25], the IMT predicted by PBE shifts toward lower pressures, away from the experimental data, for example by more than 50 GPa below the transition observed by Zaghoo and Silvera [10] based on reflectivity. Incorporation of van der Waals (vdW) interactions by Knudson [13] through the use of the vdW-DF1 [26] and vdW-DF2 [27] XC functionals shifts the predicted IMT boundary substantially, by 100 and 200 GPa, respectively. So far as we know, no DFT calculation of the H IMT prior to this report has used a high-quality meta-GGA XC functional.

Multiple predictions of the hydrogen IMT [28,29] also have come from coupled electron–ion Monte Carlo (CEIMC) calculations [30], as well as from other quantum Monte Carlo (QMC) methods [1]. In principle, CEIMC is an alternative to DFT that can treat the electrons and ions on the same footing. CEIMC calculations in practice however are a combination of methods. Ground-state QMC (not finite-T) is used to generate the electronic forces that drive the ions in the Born-Oppenheimer (BO) approximation [30]. Ionic position snapshots then are used to generate Kohn-Sham (KS) orbitals and eigenvalues from the HSE XC [31] functional and those in turn are used to calculate Kubo-Greenwood (KG) conductivities [32,33]. Obviously the ground-state QMC electron density that drives the ions has no systematic relationship to the finite-T DFT electron density that determines the electrical conductivity. An underlying difficulty is that the starting point for the electronic QMC calculations is a set of KS orbitals. Different approximate XC choices for generating them are reported to shift the calculated IMT by as much as 25 GPa (see supplemental material of reference [28]). The effect of using HSE-derived orbitals has not been investigated. The choice of the HSE range-separated hybrid functional itself is arbitrary, albeit physically defensible. By introducing exact exchange screened to short range, the spectra of such functionals mimic the spectra of the proper self-energy operator.

With a collection of methods comprising CEIMC in practice, it is important that those results be validated by a consistent methodology, particularly since predictions based on alternative QMC methodology predict a boundary that is in excess of 50 GPa lower in pressure [1]. The implicit, unanswered question is this: what IMT emerges from a straight-forward, internally consistent DFT calculation with a modern semi-local meta-GGA functional?

We address that question by just such a set of calculations of the hydrogen IMT boundary . We use the most refined meta-GGA extant, SCAN [34]. It is constructed to satisfy all 17 known constraints that a meta-GGA can satisfy and is normalized to be exact or nearly exact on systems for which an approximate XC functional can be exact. It also has been demonstrated to perform well on both molecules and solids and to work well in combination with the rVV10 [35,36] long-range vdW functional [37]. SCAN + rVV10 thus is an exemplary candidate to address the IMT boundary issue. With SCAN + rVV10, we provide a complete, consistent DFT treatment of that issue by determining the effects of generalized Kohn-Sham (gKS) versus ordinary KS treatments and by delineating the extent of NQEs via comparison of ab initio path integral molecular dynamics (quantum nuclei, PIMD) and BO molecular dynamics (classical point nuclei, BOMD) calculations.

*Method and procedures.* BOMD calculations were done with the Vienna ab initio Simulations Package (VASP) [38-40]. All simulations used a system of 256 atoms in a cubic supercell, lattice constant ranging from 7.05 to 10.27 Å, at the Γ point with 256 bands. The bare Coulomb ionic potential was handled with the projector augmented wave (PAW) [41] method. (Note that these are PBE PAWs, as is the case with all other SCAN calculations in VASP that use PAWs.) Initial convergence tests were performed at 0.6 g/cm$^3$ and 2500 K to ensure the energy and pressure converged within 1 and 2.5% respectively, with subsequent checks performed throughout the study. Ring-polymer PIMD [42] calculations were done with the i-PI interface [43,44] with Quantum-Espresso [45,46] using 8 beads. All PIMD and BOMD simulation parameters were consistent. However the PIMD calculations used a local pseudo-potential [47]. It is important to note that we did PIMD simulations for both hydrogen and deuterium, a distinction not made previously (DFT or CEIMC).

Total computational costs from use of a meta-GGA XC functional can be formidable. Therefore, for the majority of the BOMD simulations, we used the de-orbitalized version of SCAN, SCAN-L [48,49]. Similarly, the majority of the PIMD simulations used SCAN because SCAN-L was not available in Quantum-Espresso at the time. In both simulation types the rVV10 correction was used. From 20 snapshots for each trajectory we did KG calculations of the dc conductivity, extracted in the static field limit, and of the reflectivity; as well as analysis of the ionic pair correlation functions (PCF) and band gaps. For these, the KG calculations used KS eigenvalues and orbitals from SCAN-L + rVV10.

Several subsidiary calculations confirm the limited, inconsequential effects of SCAN versus SCAN-L. KG dc conductivity calculations were done with both SCAN + rVV10 and SCAN-L + rVV10 orbitals and eigenvalues on the same sets of PIMD SCAN + rVV10 ion configurations. It should be noted the SCAN + rVV10 KG calculations are performed in QE, see reference [50] for details, while the SCAN-L + rVV10 KG calculations are performed in VASP. Differences in ion configurations from SCAN versus SCAN-L were assessed by doing BOMD calculations with SCAN along two isochores, 0.8 and 1.0 $g/cm^3$. Details of these results and a complete list of simulation details can be found in the supplemental material [51].

To follow a thermodynamic path consistent as closely as possible with the SCEs [7-12] the MD simulations were done along sets of isochores (with the exception of the two lowest pressure transition points which followed the 2500 and 3000 K isotherms) with densities ranging from 0.8 to 1.15 $g/cm^3$. Consistent with the literature [6,13,14], a minimum dc conductivity of 2000 S/cm, albeit an arbitrary value see discussion in reference [10], was the primary criterion for determining the location of the metallic transition.

*Results* Figure 2(a) shows the dc conductivity for H at 0.9 and 1.0 $g/cm^3$ from both PIMD with SCAN ionic configurations and BOMD with SCAN-L configurations. Measured against the 2000 S/cm criterion, there are substantial increases of four orders of magnitude in the dc conductivity for both the classical and quantum nuclei systems. The corresponding reflectivity, Fig 2(b), at ≈1.96 eV (calculated relative to vacuum), consistent with the measurements of Zaghoo and Silvera [10], tends to have an abrupt slope change (see supplemental material [51]) that coincides with the dc conductivity increase, consistent with an IMT. Observe however, that unlike the experimental observation [10,12] of reflectivity saturating at 0.55, Figure 2b indicates a possible saturation around 0.65; a feature that appears to be dependent on the thermodynamic conditions of the system [51].

The ionic PCF provides an upper bound on the number of possible molecules [52]. Analysis of it, Fig 2(c), indicates a rapid change in the molecular character of the system at the metallic transition onset. Inclusion of NQEs dramatically enhances $H_2$ dissociation prior to the onset of metallic character. This is indicated by a clear slope change, compared to the classical ion case, in the height of the first PCF peak at the 2000 S/cm criterion. That increase in the cumulative amount of dissociation shifts the resulting IMT towards lower temperatures (recall Figure 1) and produces a steeper slope in both the dc conductivity and reflectivity of the system as compared to the classical ion case. This connection between the dissociation and metallization has been demonstrated experimentally [53] at pressures and temperatures just outside the range investigated in this report.

Inclusion of NQEs also induces a clear splitting of the IMT boundaries of hydrogen and deuterium. That splitting is an unambiguous isotope effect as only the ionic mass differs between the two PIMD simulations. Although the splitting is apparent, it is ≈ 300 K smaller than what is observed experimentally [10,12]. This suggests that NQEs alone, as incorporated in this context, may not be the sole cause of the IMT isotopic boundary splitting.

Figure 2(d) is an analysis of the indirect energy gap of the system. We define that gap in terms of the highest occupied and lowest unoccupied KS bands at T = 0 K (HOMO, LUMO respectively). They can be identified by symmetry at any temperature. (Observe that the HOMO-LUMO gap defined in this way is a good approximation for the finite temperature system of interest here because the relevant temperatures never exceed two percent of the electron Fermi temperature. This is the same reason that the ground state approximation is valid for the XC free energy; recall discussion above.) The bands in this case are from SCAN-L. As may be seen, in both the classical and quantum ion cases there is a clear energy gap closure that lies within 100 K of the onset of the minimum metallic dc conductivity. This further justifies the use of 2000 S/cm as a stand in for the primary criterion marking the onset of the metallic transition in this report. This energy closure together with the PCF analysis suggests strongly that the metallic transition of fluid hydrogen is driven by abrupt band gap closure associated with the dissociation of insulating molecular hydrogen into metallic atomic hydrogen.

Returning now to the resulting transition boundary of Figure 1, there are several features worth noting. First, the resulting IMT boundary calculated here with NQEs is in good agreement with the experimental data for hydrogen, lying as it does between the absorption onset [7-9,11] and the reflectivity increase [10]. This feature is important, because all conventional semi-local functionals, including SCAN-L, underestimate the insulating band gap [54], hence might lead to overestimated conductivity and underestimation of the IMT temperature. We have characterized the magnitude of that effect by KG calculations with both SCAN + rVV10 and SCAN-L + rVV10 to determine the dc conductivity on the same set of ion configurations (from the PIMD SCAN + rVV10 trajectories). The average difference between the two IMT boundaries (again with the 2000 S/cm criterion) is 11 K. At no point does the difference exceed 20 K. In contrast, the shift in the metallic transition temperature associated with the standard error on the dc conductivity, from the set of snapshots, is ± 30 K. Further discussion is in the Supplemental Material [51].

Second, as evident from this work and previous studies [4,13,18,19,20,49], the underlying set of ion configurations at each set of thermodynamic conditions has the strongest influence on the resulting IMT. Thus it is important to know the magnitude of effects of trajectory differences between SCAN + rVV10 and SCAN-L + rVV10. For this assessment, we compare the dc conductivity obtained from KG with SCAN-L + rVV10 based on snapshots from BOMD with SCAN-L + rVV10 versus BOMD SCAN + rVV10 with KG SCAN-L + rVV10. The 0.8 and 1.0 g/cm$^3$ isochores were chosen as representative. On both isochores the shift caused by changing from SCAN to SCAN-L is 5% of the IMT temperature. Again, see the Supp. Material for further detail [51].

Third, the inclusion of the long-range vdW interactions with the rVV10 correction appears to have the largest impact on the IMT boundary between 100 and 200 GPa. Inclusion tends to lower this boundary position by ≈ 150 K, with little to no effect outside of this pressure range. Fourth, in the low pressure, high temperature regime there is evidence of a convergence of the SCAN-L + rVV10 and the PBE [18] IMT boundaries that is further supported by the calculated optical and structural properties (see Supplemental Material [51]). This suggests that above such conditions, the use of PBE will provide about the same accuracy as the SCAN meta-GGA XC functional. That similarity has important implications for existing intra-planetary models that rely on the use of PBE.

In summary, we have reexamined the problem of determining the IMT boundary of warm dense fluid hydrogen with consistent use of what is arguably the best approximate XC functional currently available for treating both molecular and condensed phase systems even-handedly. The resulting hydrogen IMT boundary is in good agreement with experimental measurements across a wide range of pressures and temperatures. Analysis of the optical and structural properties shows concurrent, abrupt changes at the onset of a minimum metallic behavior. This signals a metallic transition due to an abrupt energy gap closure associated with the dissociation of molecular to atomic hydrogen. Furthermore, the inclusion of NQEs produces an explicit isotope effect in the form of clear splitting in the hydrogen and deuterium IMT boundaries.

**Acknowledgements:** This work was supported by the Department of Energy National Nuclear Security Administration Award Number DE-NA0003856 and US National Science Foundation PHY Grant No. 1802964. DMR and SBT acknowledge support by U.S. Dept. of Energy grant DE-SC 0002139. All computations were performed on the LLE HPC systems.

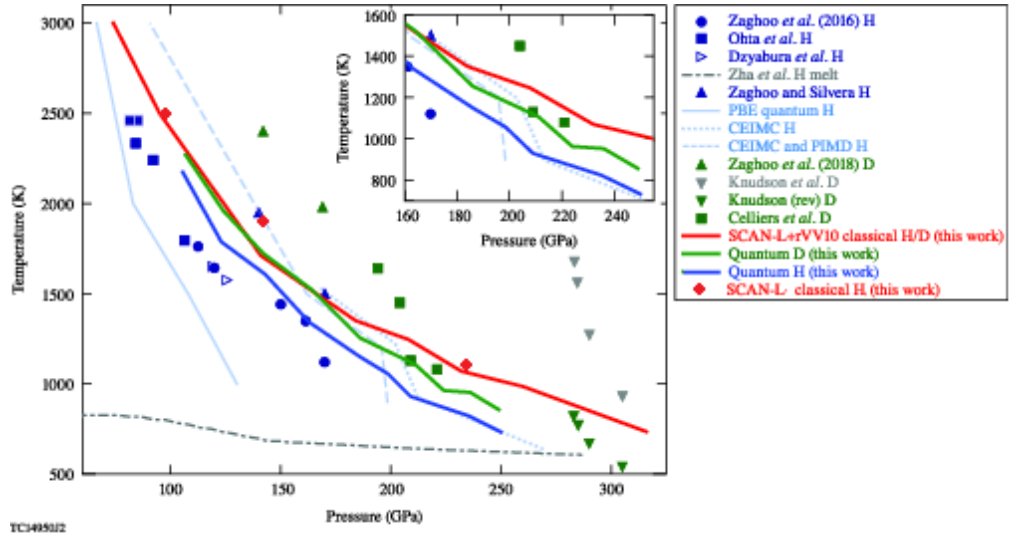

Fig. 1. Emerging picture of the hydrogen and deuterium insulator-to-metal transition (IMT). Blue symbols show experimental results for hydrogen [7-10]. Upright blue triangles are the IMT from measured reflectivities [10]. All others are from the laser heating curve plateau. Green symbols show reflectivity results for deuterium [12-14]. The black curve is the melt line based on Ref. [55]. All other solid IMT curves are theoretical predictions. The red curve is the SCAN-L + rVV10 prediction with classical nuclei and the red diamonds are three of the same classical nuclei predictions without the rVV10 correction. The corresponding blue and green curves are the predictions with NQEs for hydrogen and deuterium respectively. Subfigure: With NQE inclusion an apparent step in the IMT boundary appears. This feature has not been seen previously in DFT studies. Further analysis is required to ascertain the underlying cause of that feature.

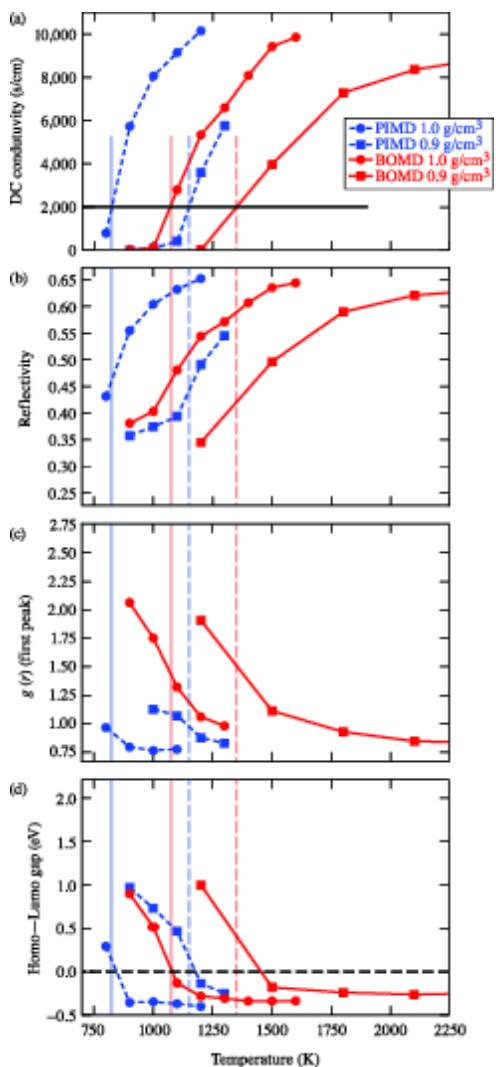

Fig 2. (a) Calculated dc conductivity of hydrogen along the 0.9 (squares) and 1.0 g/cm$^3$ (circles) isochores for quantum (blue) and classical (red) ions. The black horizontal line indicates the primary IMT location criterion: metallic dc conductivity > 2000 S/cm. Vertical lines are guides to the eye to mark the temperature at which 2000 S/cm is crossed. (b) Corresponding reflectivity at ~1.96 eV relative to vacuum. (c) Height of the first peak of the ionic PCF, a measure of the molecular character of the system. (d) Indirect energy band gap (identified from the 0 K HOMO-LUMO gap; see text). The horizontal dotted black line indicates gap closure. Negative values indicate that the LUMO minimum lies below the HOMO maximum.

Supplemental Material for

"Fully Consistent Density Functional Theory Determination of the Insulator-Metal Transition Boundary in Warm Dense Hydrogen"

Joshua Hinz[1], Valentin V. Karasiev[1]*, S. X. Hu[1], Mohamed Zaghoo[1], Daniel Mejía-Rodríguez[2], S.B. Trickey[2], and L. Calderín[3]

*vkarasev@lle.rochester.edu

[1] Laboratory for Laser Energetics, University of Rochester, Rochester, NY 14623

[2] Quantum Theory Project, Department of Physics, University of Florida, Gainesville, FL 32611

[3] Department of Materials Science and Engineering, University of Arizona, Tucson AZ 85721

06 Feb. 2020

*Ab initio* **simulations:**

Born-Oppenheimer molecular dynamic (BOMD) simulations were performed via the Vienna *ab initio* simulation package (VASP) [1-3] using the SCAN-L [4,5] exchange-correlation (XC) functional. The rVV10 [6,7] correlation correction was included in all simulations, unless specifically stated otherwise. The motive is to incorporate the long range van der Waals interaction that are not represented well by SCAN (or SCAN-L), despite its reasonable treatment in the vicinity of equilibrium bond lengths. Those long-range contributions may be important role in $H_2$ dissociation. All simulations used the NVT ensemble on a system of 256 atoms in a cubic super cell, with lattice parameters ranging from 7.05 to 10.27 Å. Electronic structures were calculated at the Γ point with 256 bands. Convergence tests, see Figure S1, indicate that use of Γ-point-only sampling introduces a maximum pressure uncertainty of 3%. Each MD trajectory consists of 5000 to 6000 steps with a time step of 0.1 fs, giving a total simulation time of 0.5 to 0.6 ps. The bare ion Coulomb potential was treated via the projector augmented wave (PAW) framework [8] with the use of PBE PAWs (metaGGA PAWs are not available in VASP at present).

Ring polymer path integral MD (PIMD) calculations for inclusion of nuclear quantum effects (NQEs) were done with the i-PI [9,10] interface to Quantum Espresso (QE) [11,12]. Those simulations used the SCAN [13] XC functional with the rVV10 van der Waals correction. These calculations used a local pseudo-potential [14] while the rest of the technical parameter values were as consistent possible with those used in the BOMD simulations. All MD simulation parameters are tabulated in Table S1 for reference.

With the exception of the 2500 and 3000 K IMT points, all other transition points were obtained from simulations along various isochores with densities ranging from 0.8 to 1.15 g/cm$^3$. This procedure was chosen so as to follow thermodynamic paths consistent with those in the static compression experiments. As a technical aside, as we moved along the isochore from low to high temperature, we took the initial snapshot from the previous temperature point in an attempt to converge the simulation faster by avoiding over-dissociated initial configurations. Furthermore, the isotherms in the low pressure regime, which have a density range of 0.5 to 0.75 g/cm$^3$, were chosen because of the unknown steepness of the IMT boundary slope at the outset of this investigation.

**Optical calculations:**

The dynamic conductivity was calculated in the Kubo-Greenwood (KG) formalism [15,16] and the dc conductivity extracted in the static field limit. All KG calculations in the main text (BOMD and PIMD boundaries) were performed with VASP with the use of SCAN-L + rVV10 on a set of 20 evenly spaced snapshots from each MD trajectory. Note that the first 1000 steps of each trajectory were skipped prior to taking the snapshots so as to allow for equilibration of the simulation. The KG calculations used an automatically generated 2×2×2 Monkhorst-Pack *k*-mesh with a plane-wave energy cutoff of 1000 eV and 256 bands. Results of KG convergence tests for various thermodynamic conditions are shown in Figure S2. A potential inconsistency may be introduced in the calculated PIMD IMT boundary as SCAN + rVV10 is used to produce the ionic configurations and SCAN-L + rVV10 then is used to generate the Kohn-Sham orbitals and eigenvalues for the calculated conductivity at each ionic configuration taken in the set of snapshots. See discussion below regarding the small magnitude consequences of the inconsistency.

The temperatures for which the average dc conductivity is directly above or below the 2000 S/cm criterion were identified. Linear interpolation between the two points then pinpointed the transition temperature at which a dc conductivity of 2000 S/cm occurred. To estimate the uncertainty in the calculated IMT temperature, the standard error of the dc conductivity was added (or subtracted) from the dc conductivity for the two points in the fit and a reassessment of the transition temperature made. This procedure yields an estimated maximum uncertainty associated with how the IMT was calculated, of ± 30 K. However, this is not a direct measure of uncertainty from the theoretical approximations (primarily the XC approximation).

The corresponding reflectivity was calculated from the dynamical conductivity according to reference [17]. Figure S3 shows the dc conductivity and reflectivity at 1.96 eV (calculated relative to vacuum) for various isochores (some of which are not shown in the main text).

**SCAN vs. SCAN-L:**

As noted above and in the main text, the use of SCAN in the PIMD simulations and SCAN-L in the KG calculations introduces inconsistency in the calculated PIMD IMT boundaries. SCAN-L has been shown to reproduce, or nearly reproduce, various structural and energetic quantities from SCAN for solids [5] and molecules [4]. The inconsistency arises in part because of the inequivalence of the Kohn-Sham (KS) and generalized Kohn-Sham (gKS) schemes used to determine the ground state electronic structure [4,5]. For reasons of complexity and computational cost, SCAN calculations use gKS. SCAN-L calculations use KS. We did several calculations to quantify the effects of that difference.

First we calculated hydrogen dc conductivities (with QE KGEC [18]) from SCAN + rVV10 gKS orbitals and eigenvalues on the same set of ionic snapshots from the PIMD trajectories that we used with SCAN-L + rVV10 KS orbitals and eigenvalues in the KG calculations with VASP. The resulting transition temperature had an average increase of 11 K. *None* of the transition points had a shift greater than 20 K. This shift is smaller than the estimated methodological uncertainty; recall discussion above. As such the difference between SCAN and SCAN-L (and the corresponding pseudo-potentials) inputs to the KG calculations of the PIMD boundary is inconsequential.

The second issue regarding the consistency of SCAN versus SCAN-L is the matter of the ionic trajectories and snapshots. That is, the comparison of PIMD and BOMD IMT boundaries includes both the direct electronic structure distinctions between SCAN and SCAN-L but also the possible direct NQEs. For a transition boundary that is clearly dependent on the ionic configuration set, even minor changes in its generation might have a significant impact on the IMT temperature. We quantify that effect next.

Figure S4 shows a direct comparison of the ionic pair correlation function (PCF) from SCAN-L + rVV10 and SCAN + rVV10 hydrogen BOMD simulations at 1.0 g/cm$^3$ and 900 K. It is clear that SCAN produces a system with a smaller molecular character (height of the first peak) and molecules with a longer equilibrium bond length (position of first peak). Further calculations of the dc conductivity along the 0.8, 0.9 and 1.0 g/cm$^3$ isochores for BOMD simulations are shown in Figure S5. Going from SCAN-L to SCAN to drive the BOMD simulation shifts the IMT boundary higher in temperature by 5 to 7% for all three isochores. Thus the shift in the IMT boundary associated with the inclusion of NQEs is underestimated by roughly 100 K (effective cancellation of the two changes) when comparing the SCAN-L + rVV10 BOMD results to the PIMD SCAN + rVV10 results.

**PBE comparison 3000 K:**

As can be seen in Figure 1 of the main text, the predicted IMT boundary from SCAN-L + rVV10 (BOMD) appears to approach that from PBE around 3000 K. To investigate this behavior further, the dc conductivities and the height of the first peak of the PCF from PBE and SCAN-L+rVV10 (both classical nuclei) along the 2500 K and 3000 K isotherms are compared in Figure S6. At 3000 K, one sees clearly that the difference between PBE and SCAN-L + rVV10 results for both properties becomes significantly smaller than at lower temperatures across the whole density (or pressure) range of the isotherm. Furthermore, there is a striking similarity to the results from the two functionals for the overall structure of the curves of the two properties at 3000 K that does not appear in the comparison along the 2500 K isotherms. This finding helps support the notion that the predicted IMT boundaries of SCAN-L + rVV10 and PBE do in fact become close above 3000 K and that the two XC functionals predict a fluid hydrogen system with a similar level of accuracy at such thermodynamic conditions.

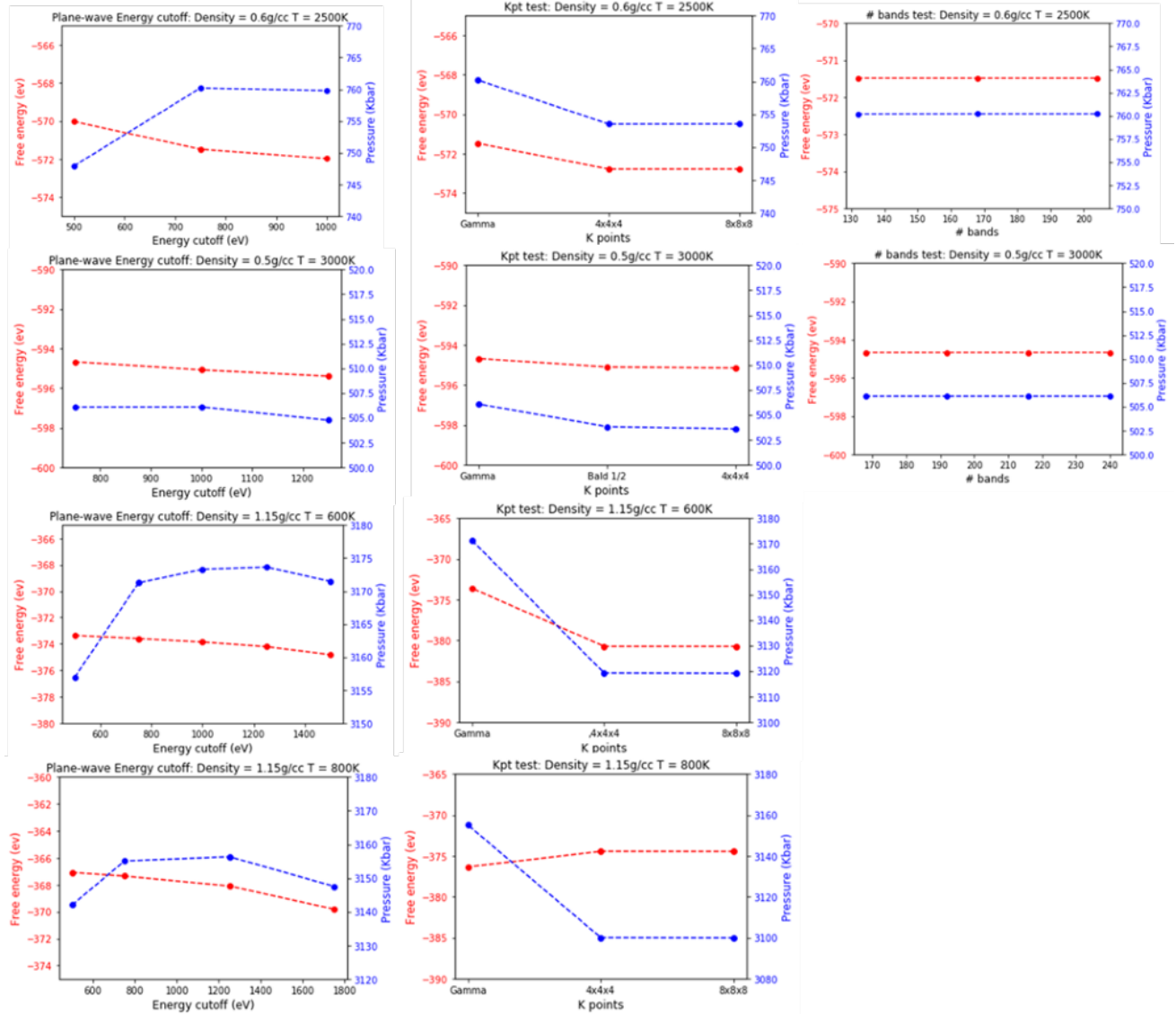

**Fig. S1.** Convergence tests performed by single point calculations for the setup of the MD simulation parameters including the plane –wave energy cutoff, k-mesh and number of bands. All red curves correspond to the free energy of the system (not including contribution from the ionic entropy) and blue curves correspond to the total pressure.

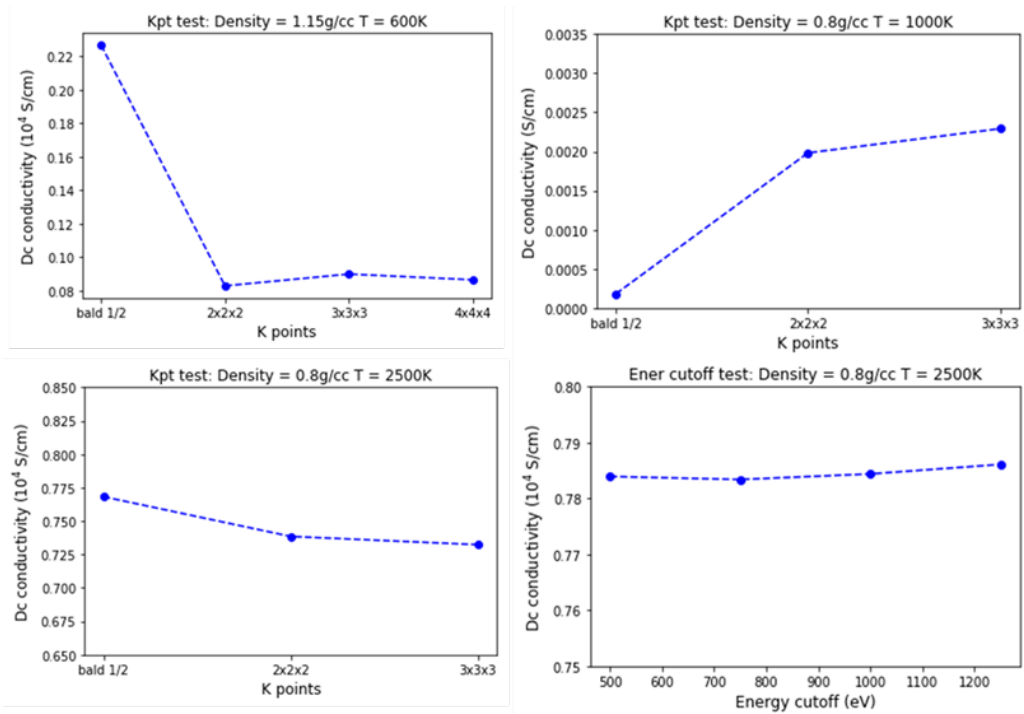

**Fig S2.** Single snapshot calculations of the dc conductivity as a function of k points (top left and right and bottom left) and plane-wave energy cutoff (bottom right) used to determine the converged set of simulation parameters for the dc conductivities calculations over the set of snapshots for each thermodynamic condition.

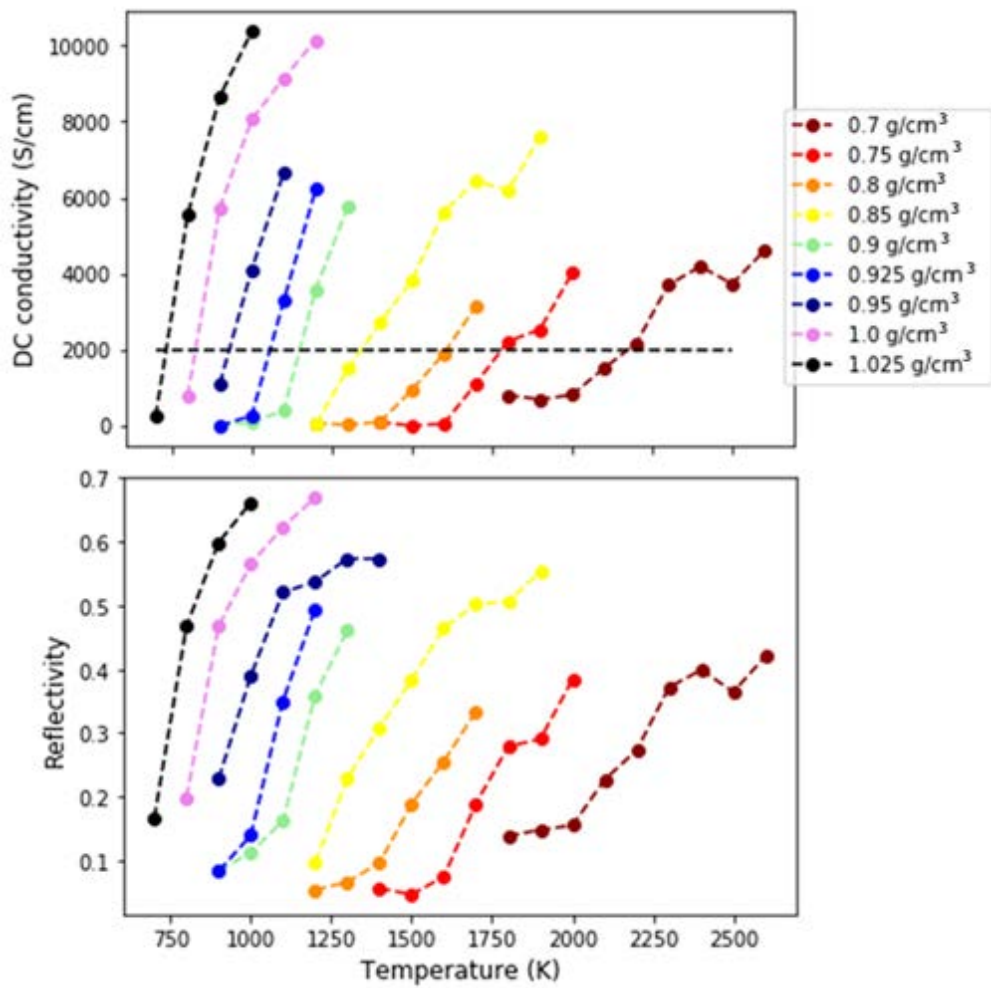

**Fig S3.** The dc conductivity (top) and reflectivity (bottom) of the PIMD isochores with the use of SCAN-L + rVV10 in the KG calculation. Note the horizontal dotted black line in the dc conductivity panel indicates the 2000 S/cm criterion used to determine the IMT.

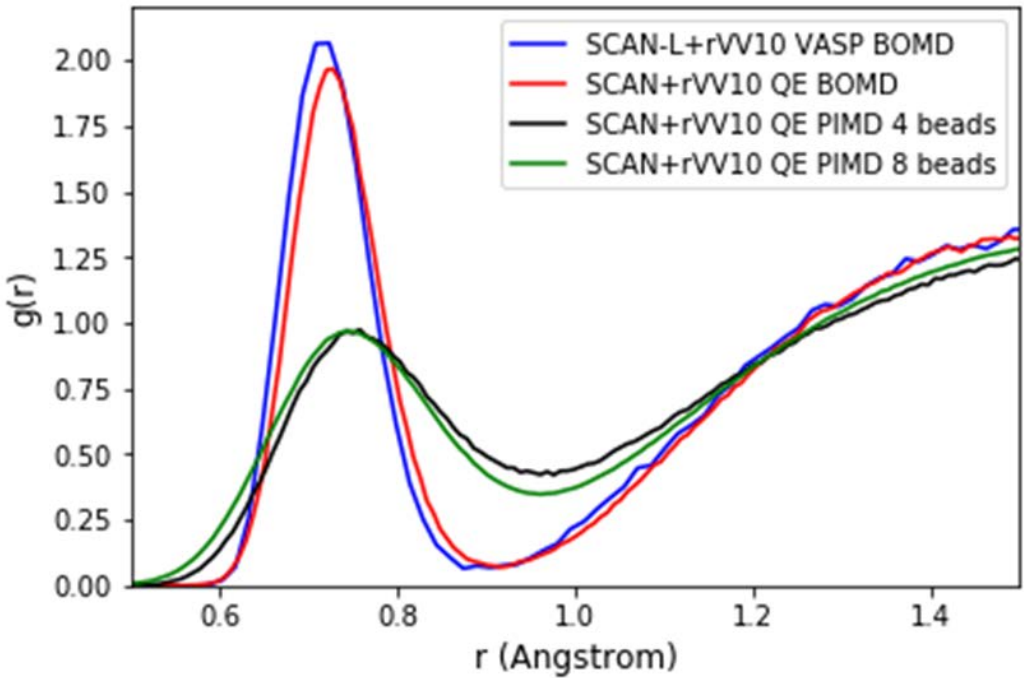

**Fig. S4.** Pair correlation function (PCF) for 4 separate simulations of hydrogen at 1.0 g/cm$^3$ and 900 K. Blue curve corresponds to the BOMD simulation with VASP using SCAN-L + rVV10. Red curve is the PCF from PIMD simulation with one bead (i.e. BOMD) using SCAN + rVV10. Green and black are PCFs from PIMD simulations with eight and four beads respectively.

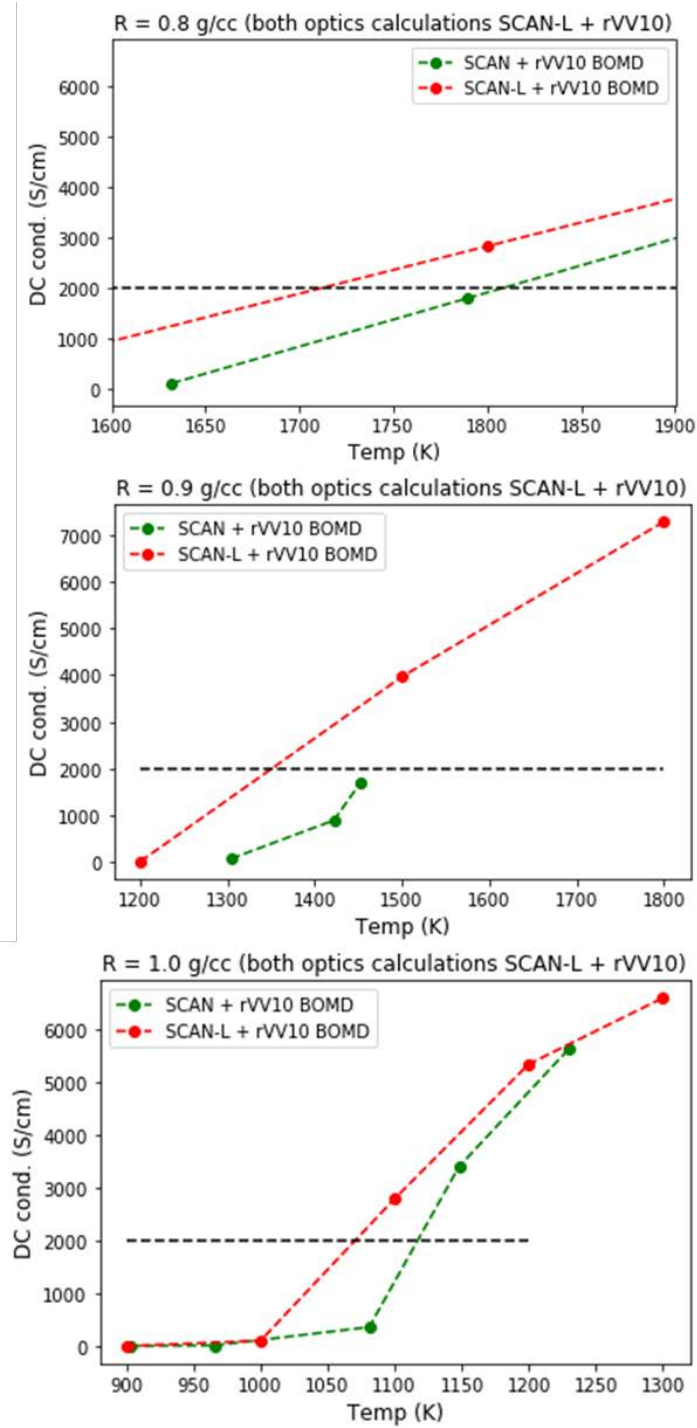

**Fig. S5.** Comparison of dc conductivities along the 0.8 (top), 0.9 (middle) and 1.0 g/cm$^3$ (bottom) isotherms. The dotted black line indicates a dc conductivity of 2000 S/cm. Each red curve corresponds to the BOMD simulation with the use of SCAN-L + rVV10 and the green curve corresponds to BOMD simulations with SCAN + rVV10. Note that the KG calculations for both sets of data were performed with SCAN-L + rVV10.

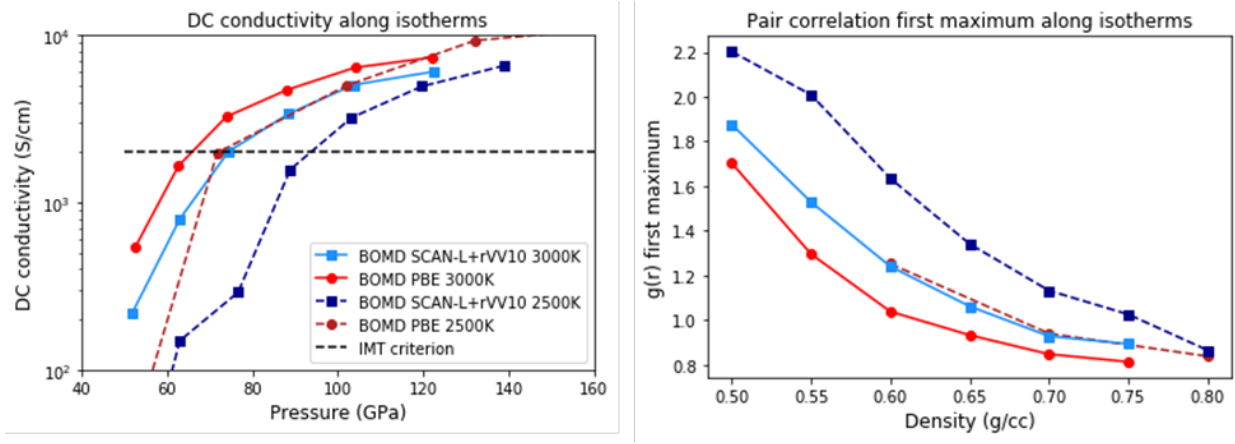

**Fig S6.** Left, comparison of the dc conductivity along the 2500K and 3000K isotherms for SCAN-L + rVV10 and PBE (BOMD). Right, first peak of the pair correlation function for the corresponding dc conductivity curves in the left panel.

| Parameter | BOMD | PIMD |
|---|---|---|
| # atoms | 256 | 256 |
| Plane-Wave cutoff energy | 750 eV | 2700 eV |
| Convergence criterion on the total electronic free energy | 1E-5 eV | 2.7E-5 eV |
| # bands | 168 | 160 |
| K pt's | gamma | gamma |
| Time step | 0.1 fs | 0.2 to 0.5 (adjust for temperature and density) |
| # MD steps | 6000 | 6000 |
| Electron partial occupancies | Fermi smearing | Fermi smearing |
| Bare proton treatment | Projector Augmented Wavepotentials [8] | Local [14] |
| Ensemble | NVT | NVT |
| Thermostat | Nose-Hoover (acts every 40 MD steps) | Langevin (acts every 200 fs) |
| # beads | N.A. | 8 |

**Table S1.** Quick reference for the parameters used in the BOMD and PIMD simulations. Note, in the case of Fermi smearing the smearing temperature is set to the ionic temperature to be maintained by the thermostat.